\documentclass[11pt]{article}
\usepackage{amssymb}
\def\bea{\begin{eqnarray}}
\def\eea{\end{eqnarray}}
\def\nn{\nonumber}
\def\be{\begin{equation}}
\def\ee{\end{equation}}
\def\lb{\label}

\def\be{\begin{equation}}
\newcommand{\cF}{{\cal F}}
\newcommand{\cA}{{\cal A}}

\newcommand{\e}{\mbox{\rm e}}

\newcommand{\ds}{\displaystyle}

\def\nqq{\hspace*{-2em}}
\def\lal{&&\nqq {}}
\def\beq{\begin{equation}}
\def\eeq{\end{equation}}
\def\const{{\rm const}}

\begin{document}

\begin{flushright} DTP-MSU/05-11

\end{flushright}

\begin{center}

\vspace{2cm}

{\LARGE {\bf Liouville and Toda dyonic  branes: \\regularity and BPS limit} }

\vspace{1cm} {\bf Dmitri V. Gal'tsov and Dmitri G. Orlov}

\vspace{1cm}

{\it Department of Theoretical Physics, Moscow State
University, 119899, Moscow, Russia}

\vspace{0.5cm}  e-mails: {gdmv04@mail.ru; orlov{\_}d@mail.ru}

\end{center}

\vspace{1cm}

\begin{abstract}
We reconsider dyonic p-brane solutions derivable from Liouville and
Toda integrable systems and investigate their geometric structure.
It is  shown that the  non-BPS non-black dyonic branes are not
regular on the horizon.
\end{abstract}
\vfill \eject

\section{Introduction}
Einstein equations with the antisymmetric form and   dilaton sources
describing singly charged   branes with the metric possessing
$SO(p,1)\times SO(D-p-1)$ isometries are fully integrable via
reduction  of the system to uncoupled Liouville equations
\cite{Lu1996:la,2,ChGaGu02}.  For an even dimension of the
space-time the similar metric ansatz leads to Liouville or Toda
integrable systems also in the case of {\it dyonic branes}
\cite{Lu1996:la}, now for certain discrete values of the dilaton
coupling constant. Some subclasses of these solutions are also
encountered within the case of the intersecting branes
\cite{1,2,14,141,15,16,3,4,10}.

Integration of the system leads to a generic solution containing the
number of integration constants. It was claimed in the literature
that extra parameters other than charges and the event horizon
radius may be associated with additional physical structures such as
tachyon on the brane \cite{14}. Meanwhile, the detailed
investigation of the geometric structure of the solution in the case
of a single charge revealed that the additional parameters lead to
naked singularities \cite{GaLeCl04,Gal'tsov:2005vf}; this does not
invalidate these solutions, but raises the question of resolution of
singularities in more general context. The purpose of the present
paper is to investigate the singularity structure of  dyonic branes.
Note that branes with both electric and magnetic charges may exist
in any space-time with electric and magnetic branes having different
dimensions (branes within branes of the type of Ref. \cite{5}. In
even dimensions and with the antisymmetric form of a suitable rank,
both electric and magnetic branes may have the same dimensions
\cite{6}. Here we will be interested by dyonic branes of this latter
type, which are derivable from Liouville and Toda systems. Due to
complete integrability of the equations one is able to obtain the
solution with the maximal number of free parameters and then to
determine the values of parameters suitable for asymptotically flat
solutions without naked singularities. In particular, we reproduce
the solutions of the Ref. \cite{Lu1996:la} and reveal the presence
of naked singularities in the case of non-BPS non-black branes.

\section{General setting}

We consider the standard (Einstein frame) action describing
classical branes in supergravities which contains the metric, the
antisymmetric form $F_{[q]}$, and the dilaton  $\phi$ with the
coupling constant $a$:
\bea
\label{action}\!\!\!\!\!\!\!\!\!
S = \int d^d x \sqrt{-g} \left( R - \frac12 \partial_\mu \phi
\partial^\mu \phi - \frac1{2\, q!} \, \e^{a\phi} \, F_{[q]}^2
\right), \eea the corresponding equations of motion being \bea
R_{\mu\nu} - \frac12 \partial_\mu \phi \partial_\nu \phi-\frac{\e^{a\phi}}{2(q-1)!}\lal\nn\\
\times\left[F_{\mu\nu_2\cdots\nu_q}F_\nu{}^{\nu_2\cdots\nu_q}- \frac{q-1}{q(d-2)}F_{[q]}^2\,g_{\mu\nu}\right]\lal=0,\label{Ein} \\
\partial_\mu \left( \sqrt{-g} \, \e^{a\phi} \,F^{\mu\nu_2\cdots\nu_q} \right)\lal=0, \label{form} \\
\frac1{\sqrt{-g}}\, \partial_\mu \left( \sqrt{-g}\partial^\mu\phi
\right) - \frac{a}{2\, q!} \e^{a\phi}F_{[q]}^2\lal=0\label{dil}.
\eea

Consider the space-time consisting of the ${(p+1)}$-dimensional
brane world-volume and  the transverse ${q=D-p-1}$-dimensional
space $\Sigma_{k,\sigma}\times\mathbb{R}_{q-k}$:
\bea\label{metric}
ds^2 = - \e^{2 B}dt^2 + \e^{2 D} (dx_1^2 + \ldots + dx_p^2)+\e^{2 A}dr^2+\e^{2 C} \, d\Sigma_{k,\sigma}^2 + \e^{2 E}(dz_1^2
+ \ldots +dz_{q-k}^2),
\eea
where five metric functions $A(r),\,B(r),\,C(r),\,D(r)$ and $E(r)$
depend on the single radial variable $r$. Here
 $\Sigma_{k,\sigma}$ with $\sigma=0,+1,-1$ is the $k$-dimensional
 flat,
 spherical or hyperbolic space with isometries ${SO(k-1,1)},\, {ISO(k)}$ and
$SO(k)$ respectively:
\bea
d\Sigma_{k,\sigma}^2 = \bar g_{ab} dy^ady^b= \left\{
 \begin{array}{ll}
 d \varphi^2 + \sinh^2\varphi \, d\Omega_{(k-1)}^2, \qquad & \sigma=-1,\\
 d \varphi^2 + \varphi^2 \, d\Omega_{(k-1)}^2, \qquad & \sigma=0,\\
 d \varphi^2 + \sin^2\varphi \, d\Omega_{(k-1)}^2, \qquad & \sigma=+1,
 \end{array} \right.
\label{gmetric} \eea The corresponding Ricci tensor reads \bea \bar
R_{ab} = \sigma (k-1) \bar g_{ab}. \eea The reparametrization of the
radial variable  allows to choose arbitrarily the  gauge function
$\cF$: \bea
 \lb{gauge0}
\ln \cF=-A+B+kC+pD+(q-k)E, \eea while the equations of motion can
be solved for any
 $\cF$. Using this notation, one can present the Ricci tensor for the metric
(\ref{metric}) as follows:
\bea
\lal R_{tt}=\e^{2B-2A}\left[B''+B'(\ln\cF)'\right],\nn\\
\lal R_{\alpha\beta}=-\e^{2D-2A}\left[ D'' + D'(\ln\cF)'\right]\delta_{\alpha\beta},\nn\\
\lal R_{rr}= -B''-B'(B'-A')-k(C''+C'^2-A'C')\nn\\
\lal \!\!-(q-k)(E''+E'^2-A'E')-p(D''+D'^2-A'D'),\nn\\
\lal R_{ab}=-\left(\e^{2C-2A}\left[C''+C'(\ln\cF)'\right]-\sigma(k-1)\right)\,\bar g_{ab},\nn\\
\lal R_{ij}=-\e^{2E-2A}\left[E''+E'(\ln\cF)'\right]\delta_{ij},
\eea
where prime denotes the derivative with respect to  $r$.

Consider now the even-dimensional space-time ${d=2n}$ and assume the
rank of the form to be ${q=n}$, so that ${p=n-2}$. Then the
equations for the form field can be solved dyonically as follows:
\bea F_{[n]}=b_1{\rm vol_n}+b_2\e^{-a\phi}\ast {\rm vol_n}, \eea
where \bea {\rm vol_n}={\rm vol}(\Sigma_{k,\sigma})\wedge
dz_1\wedge...\wedge dz_{n-k}.\nn \eea
 The form is self-dual with respect to
S-duality trans\-for\-ma\-tion with magnetic and electric charges
  $b_1,\;b_2$ interchanged.

\section{Liouville case\label{afrs}}
Introducing a new independent variable
 $\tau$ via\bea \label{tau} \frac{d\tau}{dr}=\frac{(k-1)}{\cF},
\eea and denoting the derivative with respect to $\tau$ by a dot,
we obtain the following system of equations:
\bea
\lal \ddot{B}=\frac{b_1^2\e^{G_1}+b_2^2\e^{G_2}}{4(k-1)^2},  \label{eqBd}\\
\lal \ddot{D}=\frac{b_1^2\e^{G_1}+b_2^2\e^{G_2}}{4(k-1)^2},  \\
\lal \ddot{E}=-\frac{b_1^2\e^{G_1}+b_2^2\e^{G_2}}{4(k-1)^2},  \\
\lal \ddot{C}=-\frac{b_1^2\e^{G_1}+b_2^2\e^{G_2}}{4(k-1)^2}+\frac{\sigma}{k-1}\e^{2({\cal A}-C)},\\
\lal \ddot{\phi}=\frac{a}{2(k-1)^2}\left[b_1^2\e^{G_1}-b_2^2\e^{G_2}\right],
\eea
where
\bea
G_{1,2}=\pm a\phi+2B+2(n-2)D,\nn
\eea
together with the constraint
\bea
\label{ceqd}
-\dot{\cal A}^2+\dot{B}^2+k\dot{C}^2+(n-2)\dot{D}^2+(n-k)\dot{E}^2+\frac12\dot{\phi}^2=\frac{b_1^2\e^{G_1}+b_2^2\e^{G_2}}{2(k-1)^2}-\sigma \frac{k}{k-1}\e^{2({\cal A}-C)}.
\eea
Here ${\cal A} =A+ \cF$. For the linear combination of the metric
functions $H=2({\cal A}-C)$ one then finds the separated equation:
\bea
 \ddot{H}=2\sigma \e^{H},
\eea which has a solution:
\bea
\lal H=\left\{
\begin{array}{ll}
\ln\left[\frac{\beta^2}4\right]-\ln\left[\sinh^2\left(\frac{\beta}2(\tau-\bar{\tau})\right)\right],&\sigma=1, \\
\beta(\tau-\tau_1),&\sigma=0, \\
\ln\left[\frac{\beta^2}4\right]-\ln\left[\cosh^2\left(\frac{\beta}2(\tau-\bar{\tau})\right)\right],&\sigma=-1.
\end{array}\right.\nn\\
\lal
\eea The functions $G_1$ and $G_2$ satisfy the equations:
\bea
\lal \ddot{G_1}=\frac{b_1^2\Delta_1\e^{G_1}+b_2^2\Delta_2\e^{G_2}}{2(k-1)^2}, \\
\lal \ddot{G_2}=\frac{b_1^2\Delta_2\e^{G_1}+b_2^2\Delta_1\e^{G_2}}{2(k-1)^2},
\eea
with \bea \Delta_1=a^2+(n-1),\qquad\Delta_2=-a^2+(n-1). \nn \eea
For arbitrary values of the coupling parameter, the system for
$G_1$ and $G_2$ can not be separated. However, in two particular
cases it  reduces to separate equations. Namely, for $a=0$ one
obtains the coinciding equations
\bea
\lal  G_1=G_2=G,\qquad \Delta_1=\Delta_2=\Delta=n-1,\nn \\
\lal \ddot{G}=\frac{(b_1^2+b_2^2)\Delta \e^G}{2(k-1)^2},
\eea
the solution being:
\bea
G=\ln\left[\frac{\alpha^2(k-1)^2}{\Delta(b_1^2+b_2^2)}/\sinh^2
\left(\frac{\alpha}2(\tau-\tau_0)\right)\right]. \eea
In another
particular case $a^2=n-1$ one has $\Delta_2=0,\,\Delta_1=2(n-1)$
and  the equations for $G_1$ and $G_2$ split
\bea
\ddot{G}_{1,2}=\frac{b_{1,2}^2\Delta_1}{2(k-1)^2}\e^{G_{1,2}},
\eea
the solutions being
\bea
G_{1,2}=\ln\left[\frac{\alpha^2(k-1)^2}{\Delta_1b_{1,2}^2}/\sinh^2
\left(\frac{\alpha_{1,2}}2(\tau-\tau_{1,2})\right)\right].
\eea
Changing the notation for the case $a=0$ as follows
$b_{1,2}^2=\frac{b_1^2+b_2^2}2,\;\tau_{1,2}=\tau_{0},\;
\alpha_{1,2}=\alpha$, one can present the solution in both cases
in a unique way
\bea
G_{1,2}=\ln\left[\frac{\alpha^2(k-1)^2}{2(n-1)b^2_{1,2}}/
\sinh^2\left(\frac{\alpha_{1,2}}2(\tau-\tau_{1,2})\right)\right].
\eea

One has to consider also the case of zero $\alpha$ and $\beta$.
Taking into account the first integrals of the Liouville equations
\bea
\dot{G}_{1,2}^2-\frac{b_{1,2}^2(n-1)}{2(k-1)^2}\e^G
=\alpha_{1,2}^2,\quad\dot{H}^2-4\sigma \e^H=\beta^2, \eea one
finds:
\bea \label{gh0}
\lal G_{1,2}=\ln\left[\frac{2(k-1)^2}{b_{1,2}^2(n-1)(\tau-\tau_{1,2})^2}\right],\nn\\
\lal H=\left\{\begin{array}{cl}
H_0,&\quad \sigma=0, \\
\ln\left[\frac1{\sigma(\tau-\bar{\tau})^2}\right], & \quad
\sigma=-1,1.
\end{array}\right.
\eea
The constant $\bar{\tau}$ can be set zero since the system is
autonomous in terms of $\tau$.

For both particular cases of $a$, the metric functions and the
dilaton will be expressed through $H,\;G_1,\;G_2$, as follows:
\bea\label{lsolt}
B\lal =\frac{G_1+G_2}{4(n-1)}-\frac{n-2}{n-1}(d_1\tau+d_0),\!\\
D\lal =\frac{G_1+G_2}{4(n-1)}+\frac1{n-1}(d_1\tau+d_0),\!\\
E\lal =-B+e_1\tau+e_0=-\frac{G_1+G_2}{4(n-1)}+\epsilon_1\tau+\epsilon_0,\!\! \\
{\cal A}\lal =\frac{kH}{2(k-1)}-\frac{G_1+G_2}{4(n-1)}+c_1\tau+c_0,\\
C\lal =\frac{H}{2(k-1)}-\frac{G_1+G_2}{4(n-1)}+c_1\tau+c_0,\\
\phi\lal =\frac{G_1-G_2}{2a}=\frac1a\ln\left|\frac{b_2\sinh\left(\frac{\alpha_2}2(\tau-\tau_{2})\right)}
{b_1\sinh\left(\frac{\alpha_1}2(\tau-\tau_{1})\right)}\right|,
\eea
with free parameters $d_{0,1},\; e_{0,1}$ and \beq
\epsilon_{0,1}=\frac{n-2}{n-1}d_{0,1}+e_{0,1},\qquad
c_{0,1}=-\frac{n-k}{k-1} \epsilon_{0,1}.\eeq
 From the constraint equation one finds the
following relation between the parameters :
\bea\label{ceqdp}
-\frac{k}{4(k-1)}\beta^2+\frac1{4(k-1)}(\alpha_1^2+\alpha_2^2)+(n-k)\epsilon_1^2+(k-1)c_1^2+\frac{n-2}{n-1}d_1^2=0.
\eea

To reveal the location of singularities it is convenient first to
analyze the   scalar curvature. Using the equations of motion one
can present it as follows \bea\label{Rs}
R=\frac{(k-1)^2}{2}\e^{-2{\cal A}}\dot{\phi}^2. \eea Our solution
(\ref{lsolt}) this reduces to
\bea
\lal R=\left(\frac{k-1}{\sqrt{8}a}\right)^2
\left(\frac{4}{\beta^2}\right)^{\frac{k}{k-1}}\left[\frac{\alpha^2(k-1)^2}{2(n-1)b_1b_2}\right]^{\frac1{n-1}}
\sinh^{\frac{2k}{k-1}}\left(\beta\tau/2\right)\e^{-2(c_1\tau+c_0)}\cdot\nn\\
\lal \times\left(\sinh\left(\frac{\alpha_1}2(\tau-\tau_1)\right)\sinh\left(\frac{\alpha_2}2(\tau-\tau_2)\right)\right)^{-\frac1{(n-1)}}\nn\\
\lal \times\left(\alpha_2\coth\left(\frac{\alpha_2}2(\tau-\tau_2)\right)-\alpha_1\coth\left(\frac{\alpha_1}2(\tau-\tau_1)\right)\right)^2.\nn\\
\lal
\eea
We will also need the Kretchmann scalar
\bea\label{krech}
K=\lal R_{\alpha\beta\gamma\delta}R^{\alpha\beta\gamma\delta}=4(k-1)^4\e^{-4\cA}
\left[\sum_{i}l_i(-\ddot{Y}_i-\dot{Y_i}^2+\dot\cA\dot{Y_i})^2+\frac12\sum_{i\neq j}l_il_j\dot{Y}_i^2\dot{Y}_j^2\right. \nn\\
\lal \qquad\qquad\qquad\;\left.+\frac12\sum_{i}l_i(l_i-1)\dot{Y}_i^4-\sigma\frac{k}{k-1}\dot{C}^2\e^{2\cA-2C}+\frac12\frac{k}{(k-1)^3}\e^{4\cA-4C}\right],
\eea
where for brevity  the following multicomponent notation is
introduced $Y_i=\{B,D,C,E\},\;l_i=\{1,n-2,k,n-k\}$.

Our solution has the following special points:
$\tau=\tau_{1,2},\;\pm\infty,\;0$. At the points
$\tau=\tau_{1,2}$, which locate at finite geodesic distance (we
will consider the null geodesics) from any non-special point, the
Ricci scalar diverges. These points mark the curvature
singularities.

The point $\tau=0$ locates at an infinite null geodesic distance,
and both the Ricci and Kretchmann scalars are zero there. This
point will be identified with the asymptotic infinity.

Finally, at the limiting points $\tau=\pm\infty$ one can have
(with some parameter choice) the vanishing of the metric component
$g_{tt}=0$, which is the necessary condition for existence of the
event horizon.

\subsection{Asymptotically flat solutions with a regular horizon\label{afrs}}

Consider the case of the spherical symmetry $\sigma=1$ (for other
values of $\sigma$ no asymptotically flat solutions exist). Based
on the above analysis of special points, we can interpret
 $\tau=0$ as spatial infinity, and $\tau=-\infty$ as the event
 horizon. To clarify the geometry near the horizon consider the
 radial null geodesics curves with the affine parameter $\lambda$.
From the geodesic equation one finds: \bea\label{lrt}
d\lambda=\e^{A+B}dr=\e^{{\cal A}+B}\frac{d\tau}{k-1}. \eea The
regular horizon should be geodetically traversable, so from an
analyticity argument we have to require \bea\label{horn}
\e^{2B}\sim\lambda^n, \eea with integer $n$, namely $n=1$ for a
non-degenerate and
 $n=2$ for a degenerate regular horizons.
Near $\tau=-\infty$ our solution gives \bea \e^{2B}\sim
{\exp}\left[\frac1{(n-1)}\left(\frac{\alpha}2-(n-2)d_1\right)\tau\right],
\eea so  $g_{tt}|_{\tau=-\infty}=0$ provided \bea \alpha>2(n-2)
d_1 . \eea The geodesic equation for a non-degenerate horizon
${n=1}$ leads to \bea \ds
\frac{d}{d\lambda}\e^{2B}=\e^{-(\cA+B)}\frac{d}{d\tau}\e^{2B}\rightarrow
\const, \eea and therefore
\bea \e^{\cA}\sim \e^B. \eea
Then
the first two terms in the left hand side of the constraint
equation (\ref{ceqd}) cancel, while the remaining expression
becomes positive definite. Since the right hand side vanishes,
this means vanishing of all the derivatives at the left hand side
separately. We obtain the system of equations which has the unique
solution \bea
\frac12(\alpha_1+\alpha_2)=\beta=2e_1=-2d_1,\;\alpha_1=\alpha_2=\alpha.
\eea With this, the constraint (\ref{ceqdp}) will be satisfied for
all $\tau$. Imposing now the asymptotic flatness conditions
 \bea
 B\approx 0,\;\;D\approx 0,\;\;E\approx 0,
\eea we obtain three more conditions on the parameters
\bea
\lal d_0=0,\qquad\;e_0=0,\qquad
\left(\sinh\frac{\alpha}2\tau_{1}\cdot\sinh\frac{\alpha}2\tau_{2}\right)^2=\left(\frac{\alpha^2(k-1)^2}{2b_1b_2(n-1)}\right)^2.
\eea
The dilaton at infinity will have the finite value  \bea
\phi_{\infty}=\phi_0,\quad
\left|\frac{b_2\sinh\frac{\alpha}2\tau_{2}}{b_1\sinh\frac{\alpha}2\tau_{1}}\right|=\e^{a\phi_0}.
\eea From these equations we obtain the parameters
 $\tau_1,\;\tau_2$.
We also need the asymptotic condition for the conformal factor
$\e^{2C}\sim r^2$ in terms of the "curvature" radial variable.
This can be achieved choosing the gauge function $\cal F$ as
follows:
\bea
\lal{\cal F}=r^kf_{+}f_{-},\qquad f_{\pm}=1-\frac{\xi_{\pm}}{\xi},\\
\lal \xi=r^{k-1},\qquad\xi_{\pm}=r_{\pm}^{k-1},\qquad\xi_{-}<\xi_{+}.\nn
\eea
Here two new parameters $r_{\pm}$ are introduced marking the event
horizon and the internal horizon, such that $r_+>r_-$, and
\beq\alpha=x_{+}-x_{-}.\eeq  The corresponding coordinate
transformation reads: \bea
\tau=\frac1{\alpha}\ln\frac{f_{+}}{f_{-}}. \eea In terms of this
new radial coordinate the solution will read
\bea
\lal H=\ln\left[\xi^2f_{+}f_{-}\right],\nn\\
\lal G_{1,2}=\ln\frac{f_{+}f_{-}}{f_{1,2}^2}+
\ln\left[\frac{2(k-1)^2}{b^2_{1,2}(n-1)}\xi_{1,2}^2f_{+}^{1,2}f_{-}^{1,2}\right],\\
\lal f_{1,2}=1-\frac{\xi_{1,2}}\xi,\qquad f_{\pm}^{12}=1-\frac{\xi_{\pm}}{\xi_{1,2}},\qquad \xi_{1,2}=r^{k-1}_{1,2},\nn
\eea
where $r_{1}$($r_{2}$) are the images of $\tau_{1}$($\tau_{2}$).
The interval and the dilaton exponent then will read:
\bea\label{lsolr}
ds^2=\lal\left[\frac{f_{-}^2}{f_{1}f_{2}}\right]^{\frac1{n-1}}\left(-\frac{f_{+}}{f_{-}}dt^2+d{\bf x}^2\right) \nn\\
\lal+\left[f_{-}^{2\frac{n-k}{k-1}}f_{1}f_{2}\right]^{\frac1{n-1}}\left(\frac{dr^2}{f_{+}f_{-}}+r^2d\Sigma^2_{k,1}\right)
+\left[\frac{f_{1}f_{2}}{f_{-}^2}\right]^{\frac1{n-1}}d{\bf z}^2,\\
\e^{2a\phi}=\lal \frac{f^2_{2}}{f^2_{1}}\e^{2a\phi_0},\\
\e^{2a\phi_0}=\lal \frac{b_2^2}{b_1^2}\frac{(\xi_1-\xi_{+})
(\xi_1-\xi_{-})}{(\xi_2-\xi_{+})(\xi_2-\xi_{-})}.\nn
\eea
For $a=0$ the metric coincides with that of a singly charged brane
with the parameters $ p=n-2,\;  q=n,\; \Delta=n-1 $.

The scalar curvature in the new coordinates reads
\bea
\lal R=\left(\frac{k-1}{\sqrt2
ar^k}\right)^2\left[\frac{\xi_2}{f_2}-\frac{\xi_1}{f_1}\right]^2
\left[f_-^{2\frac{n-k}{k-1}}f_1f_2\right]^{-\frac1{n-1}}f_+f_-.
\eea
One can see that the internal horizon is singular
\bea
K\sim\alpha^4\,(n-1)(n-2)^2(n-3)\,\e^{2\alpha\tau},
\eea
except for the case of the purely spherical transverse space $n=k$ for $k=2$ or 3:

For non-zero dilaton coupling, $a^2=n-1$,  the scalar curvature in the
vicinity of  $\tau_{1,2}$ behaves as
\bea R\sim[f_1f_2]^{-\frac{2n-1}{n-1}}, \eea
while the Kretchmann
scalar for both solutions
 ${a=0}$ and ${a^2=n-1}$ diverges as
\bea R_{\alpha\beta\gamma\delta}R^{\alpha\beta\gamma\delta}
\sim[f_1f_2]^{-2\frac{2n-1}{n-1}}. \eea
Note that in the case
 $a=0$  when  $\tau_1=\tau_2=\tau_0$, the divergency of the
 Kretchmann scalar
 at $\tau_0$ is twice as strong as in the case  $a^2=n-1$.
In what follows we choose in both cases the singularities to locate
inside the event horizon.

The case of the degenerate horizon corresponds to the following
behavior of the metric function and its derivative: \bea
\e^{2B}\sim\lambda^2,\qquad\e^{B-{\cal A}}\dot{B}\sim\lambda, \eea
so we have: \bea \e^{-\cal A}\dot{B}\sim O(1). \eea Therefore
either the both factors are finite (non-zero) at the horizon, or
both $\dot{B}$ and $\e^{\cal A}$ must vanish there. Assuming the
first option for $\e^{\cal A}$, we have to admit  $\dot{\cal A}=0$
at the horizon, and similarly for $\dot{B}$ (from the constraint),
thereby coming to contradiction. Then consider the second
possibility. If $\dot{B}=0$ and $\e^{B}=0$, we find
$\alpha=2(n-2)d_1=0$. From the condition $\e^{\cal A}=0$ and the
constraint (\ref{ceqd}) we obtain  $e_1=0$, and consequently
$\beta=0$. As a result, the degenerate horizon will correspond to
the following choice of parameters \bea \alpha=\beta=d_1=e_1=0,
\eea which leads (as expected) to the limit \bea r_{-}\rightarrow
r_{+}\nn \eea in the solution with the non-degenerate horizon. In
the degenerate case the metric functions ${\cal A},\;B$ according
to (\ref{gh0}) will behave as:
\bea {\cal A}\sim\mu\ln|\tau|,\qquad B\sim \nu\ln|\tau|, \eea
where
\bea \mu=\frac1{n-1}-\frac k{k-1},\qquad\nu=-\frac1{n-1}.\nn \eea From
the geodesic equation (\ref{lrt}) and the condition (\ref{horn})
we then find: \bea
d\lambda\sim|\tau|^{\mu+\nu}d\tau,\qquad\e^{2B}\sim|\tau|^{2\nu}\sim
\lambda^{\frac{2\nu}{\mu+\nu+1}}, \eea therefore, in the
degenerate case one more condition has to be satisfied \bea
\mu+1=0\;\rightarrow\;n=k, \eea thus the degenerate regular
horizon is possible only in the case of purely spherical
transverse space. It is worth noting, that contrary to the singly
charged brane, now the vanishing of the dilaton coupling constant
is not required \cite{Gal'tsov:2005vf}.

In the final solution one can shift the origin to the point
$r=r_2$ introducing a new coordinate \bea
y^{k-1}=r^{k-1}-r_2^{k-1}.\eea This reduces the number of free
parameters by one. Thus we obtain the four-parametric dyon
solution charac\-te\-rized by two charges $b_1,b_2$, the asymptotic
value of the dilaton and the radius of the event horizon $y_+$.

\section{Toda solution}
Consider an open  Toda chain described by the lagrangian \bea
\label{Teq} {\cal L}=\frac12\sum_{i=1}^n
q_i^2+\sum_{i=1}^{n-1}g_i\e^{2(q_i-q_{i+1})}. \eea Note, that the
number of potentials is less by one than the number of independent
variables. One can show that, for a particular dilaton coupling
constant, the solution  to our system can be derived from that of
an open three-dimensional Toda chain. To this end we first
introduce the three-dimensional vector \bea
    X=\{\sqrt2B,\sqrt{2(n-2)}D,\phi/\sqrt{2}\}
\eea replacing the three initial variables (note that  the
function $E$ will differ from $B$ on the linear function, while
the equation for $H$ was already separated):
\bea
{\cal L}\lal=\dot{B}^2+(n-2)\dot{D}^2+\frac{\dot{\phi}^2}{4}+\frac{b_1^2}{4(k-1)^2}\e^{G_1}+\frac{b_2^2}{4(k-1)^2}\e^{G_2}.
\eea
We introduce also two vectorial parameters
\bea
\lal\lambda_1=\{\sqrt2,\sqrt{2(n-2)},\sqrt2a\},\nn\\
\lal\lambda_2=\{\sqrt2,\sqrt{2(n-2)},-\sqrt2a\}
\eea
and rewrite the lagrangian as  \bea {\cal
L}=\frac12<\dot{X},\dot{X}>+\sum_{i=1,2}a_i\e^{<\lambda_i,X>}, \eea
where $<.,.>$ is the euclidean scalar product with the metric
$\delta_{ij}$, and \bea a_1=\frac{b_1^2}{4(k-1)^2},\qquad
a_2=\frac{b_2^2}{4(k-1)^2}. \eea The vectors $\lambda_1$,
$\lambda_2$ are not orthogonal:
\bea
\lal<\lambda_1,\lambda_1>=<\lambda_2,\lambda_2>=2(a^2+n-1),\nn\\
\lal<\lambda_1,\lambda_2>=2(-a^2+n-1),
\eea
so it is useful to introduce the orthogonal basis $\{e'_1..e'_3\}$
adding the third vector orthogonal to the first two  as follows
\bea
\lal\lambda_1=[2(a^2+n-1)]^{\frac12}e'_2,\nn\\
\lal\lambda_2=\frac{2(-a^2+n-1)}{\sqrt{2(a^2+n-1)}}e'_2+\sqrt{\frac{8a^2(n-1)}{a^2+n-1}}e'_3,\nn\\
\lal e'_1=[e'_2,e'_3],
\eea
where $[.,.]$ means the vector product. Then we get:
\bea
\lal{\cal L}=\frac12\sum_{i,j=1}^3 \delta_{ij}{\dot{X}}'^i{\dot{X}}'^j
+a_1\exp\left([2(a^2+n-1)]^{\frac12}X'^2\right)\nn\\
\lal\qquad+a_2\exp\left(\frac{2(-a^2+n-1)}{\sqrt{2(a^2+n-1)}}X'^2+
\sqrt{\frac{8a^2(n-1)}{a^2+n-1}}X'^3\right),\nn \eea
where
$X'^i$ -- are components of $X$ in the new basis $X=X'^i e'_i$.
This system encompass also the two Liouville cases
 $a^2=n-1$  and $a=0$, when the equations separate.
 Since the coordinate $X'^1$ is cyclic, one obtains
for it the linear dependence $X'^1=x_1\tau+x_0$, while the
non-trivial part of the lagrangian reads:
\bea {\cal
L}=\frac12\left(\dot{X}'^2\right)^2+\frac12\left(\dot{X}'^3\right)^2+\sum_{i=1}^{2}a_ie^{<\lambda_i,X'>}.
\eea The potential term has the form suitable for Toda open chain
representation, but to get an appropriate kinetic term one has to
introduce an additional variable
 $X'^4$, with the kinetic term  $\frac12 \dot{X'^4}^2$, satisfying
free equations of motion. It turns out that the canonical form of
the kinetic term  (\ref{Teq}) can be achieved only for a
particular coupling constant $a^2=3(n-1)$, the corresponding
transformation to a new set of variables $\{q_1,q_2,q_3\}$ being
in this case:
\bea
\lal [2(a^2+n-1)]^{\frac12}X'^2=2(q_1-q_2), \nn\\
\lal \frac{2(-a^2+n-1)}{\sqrt{2(a^2+n-1)}}X'^2+
\sqrt{\frac{8a^2(n-1)}{a^2+n-1}}X'^3=2(q_2-q_3),\nn\\
\lal X'^4=\frac1{a}(q_1+q_2+q_3).\eea  The equation for
$X'^4$ then corresponds to the free motion of the "center of mass"
of the Toda chain
 $\ddot{q}_1+\ddot{q}_2+\ddot{q}_3=0$, and finally
the lagrangian will take the form:
\bea
{\cal L}=\frac3{a^2}\left[\frac12\sum_{i=1}^{3}\dot{q}_i^2+\sum_{i=1}^2g_i\e^{2(q_i-q_{i+1})}\right],
\eea
where
\bea
g_1=\frac{(ab_1)^2}{12(k-1)^2},\qquad g_2=\frac{(ab_2)^2}{12(k-1)^2}.\nn
\eea

Following the Ref.  \cite{Gavrilov:1994mc,Olshanetsky:1979,12,13}
we can present the solution of the Toda system as follows:
\bea
\lal g_1\e^{2(q_1-q_2)}=\frac{F_{+}}{F_{-}^2},\qquad g_2\e^{2(q_2-q_3)}=\frac{F_{-sd}}{F_{+}^2}, \nn\\
\lal F_{\pm}=\frac{4}{9A_1A_2(A_1+A_2)}\left[A_1\e^{\pm(A_1+2A_2)\tau\pm B_1}\right. \nn \\
\lal \qquad\qquad\qquad\qquad\qquad\quad\left.-(A_1+A_2)\e^{\pm(A_1-A_2)\tau\mp (B_1-B_2)}+A_2\e^{\mp(2A_1+A_2)\tau\mp B_2}\right].
\eea
Here the $A_{1,2},\,B_{1,2}$ are the constant parameters, $B_1$
and $B_2$ being fully arbitrary, while $A_1$ and $A_2$ satisfy the
sign restriction: $A_1A_2>0$. For ${a^2=3(n-1)}$ one has :
\bea
X'^2\lal=\frac{q_1-q_2}{\sqrt{2(n-1)}}=
\frac{\sqrt3}{a\sqrt8}\ln\left[\frac1{g_1}
\frac{F_{+}}{F_{-}^2}\right],\nn\\
X'^3\lal=\frac{(q_1-q_2)+2(q_2-q_3)}{\sqrt2a}=
\frac1{a\sqrt{8}}\ln\left[\frac1{g_1g_2^2}
\frac1{F_{+}^3}\right],\nn\\ \lal
\eea
and the initial variables $B,\;D,\;\phi$ are related to $X'^1$,
$X'^2$, $X'^3$ via the relations
\bea
B\lal=\frac1{\sqrt{8(n-1)}}\left[X'^2+\sqrt{3}X'^3-\sqrt{4(n-2)}X'^1\right], \nn \\
D\lal=\frac1{\sqrt{8(n-1)}}\left[X'^2+\sqrt{3}X'^3+\sqrt{\frac4{(n-2)}}X'^1\right],\nn\\
\phi\lal=\sqrt{\frac32}X'^2-\frac1{\sqrt{2}}X'^3.
\eea
Substituting the above solution for $X'$ we obtain:
\bea\label{tsol}
\lal B=-\frac{\ln\left[g_1g_2F_{+}F_{-}\right]}{4(n-1)}-
\frac{n-2}{n-1}(d_1\tau+d_0), \nn \\
\lal D=-\frac{\ln\left[g_1g_2F_{+}F_{-}\right]}{4(n-1)}+
\frac1{n-1}(d_1\tau+d_0), \nn \\
\lal E=\frac{\ln\left[g_1g_2F_{+}F_{-}\right]}{4(n-1)}+
\epsilon_1\tau+\epsilon_0,\nn\\
\lal {\cal A}=\frac{kH}{2(k-1)}+
\frac{\ln\left[g_1g_2F_{+}F_{-}\right]}{4(n-1)}+c_1\tau+c_0, \nn \\
\lal C=\frac{H}{2(k-1)}+
\frac{\ln\left[g_1g_2F_{+}F_{-}\right]}{4(n-1)}+c_1\tau+c_0, \nn \\
\lal \phi=\frac1{2\sqrt{3(n-1)}}\ln\left[\frac{g_2}{g_1}
\left(\frac{F_{+}}{F_{-}}\right)^3\right],\eea where
\bea
\lal d_i=\sqrt{\frac{n-1}{2(n-2)}}x_i,\qquad \epsilon_i=\frac{n-2}{n-1}d_i+e_i,\qquad c_i=-\frac{n-k}{k-1}\epsilon_i.\nn
\eea

Now the asymptotically flat regular solution for  $a^2=3(n-1)$ can
be performed similarly to the previous Liouville cases.
Introducing a new parameter \bea
\alpha=-\frac12\left.\left(\frac{\dot{F}_{+}}{F_{+}}+
\frac{\dot{F}_{-}}{F_{-}}\right)\right|_{\tau=-\infty}, \eea from
the constraint (\ref{ceqd}) we will get the conditions on the
event horizon : \bea \dot{C}=\dot{D}=\dot{E}=\dot{\phi}=0, \eea
the first three giving already known values of parameters: \bea
|\alpha|=\beta=2e_1=-2d_1. \eea Vanishing of the the
 dilaton derivative at the horizon means:\bea
\left.\frac{\dot{F}_{+}}{F}_{+}\right|_{\tau=-\infty}=\left.
\frac{\dot{F}_{-}}{F_{-}}\right|_{\tau=-\infty}=-\alpha, \eea
implying $A_1=A_2={\displaystyle\frac{\alpha}3}$, that leads to
simplification: \bea
F_{\pm}=\frac2{\alpha^2}\left[\e^{\pm(\alpha\tau+B_1)}-2\e^{\mp(B_1-B_2)}+\e^{\mp(\alpha\tau+B_2)}\right].\nn
\eea An asymptotic flatness condition implies
\bea\label{ctinfty}
\lal x_0=0,\qquad e_0=0,\nn\\
\lal \left.F_{+}F_{-}\right|_{\tau=0}=\frac1{g_1g_2},\qquad
\left.\frac{g_2}{g_1}
\left(\frac{F_{+}}{F_{-}}\right)^3\right|_{\tau=0}=
\e^{2a\phi_{\infty}}.
\eea
From here one finds the parameters $B_1$, $B_2$.

The case of the degenerate horizon can be treated along the same
lines as before, this leads to following conditions on the
parameters: \bea \alpha=\beta=d_1=e_1=0. \eea In this case, as can
be easily seen from the Eq. (\ref{ctinfty}), an assumption of
finite $|B_{1,2}|$) immediately gives \bea B_1=B_2=0. \eea Then
the solution degenerates \bea F_+=F_-, \eea which leads to the
constant dilaton and the condition $a=0$ in contradiction to the
assumption $a^2=3(n-1)$. Therefore, for the Toda solution the
asymptotically flat extremal configuration does not exist, the
Toda dyon exists only in the black version (in this we disagree
with the Ref. \cite{Lu1996:la}).

The curvature scalar (\ref{Rs}) for the solution obtained reads
\bea
R=\lal\frac{3(k-1)^2}{8(n-1)}\left(\frac{\dot{F}_+}F_+-
\frac{\dot{F}_-}{F_-}\right)^2\left(\frac2{\beta}\sinh(\beta\tau/2)\right)^{\frac{2k}{k-1}}
\left(g_1g_2F_+F_-\right)^{-\frac1{2(n-1)}}\e^{-2c_1\tau-2c_0}.
\eea
Similarly to the Liouville case we have two singular points
defined by the equations \bea F_+|_{\tau=\tau_1}=0,\qquad
F_+|_{\tau=\tau_2}=0. \eea To satisfy the cosmic censorship
conjecture we have to locate them inside the external
horizon, thus  $\tau_1,\;\tau_2$ can not lie in
the region $(-\infty,0]$. So we have a four-parametric solution,
characterized by the values of charges $b_1,b_2$, the value of the
dilaton at infinity $\phi_0$ and the parameter $\alpha$. Contrary
to the previous cases of the coupling constant $a=0,\;a^2=n-1$,
this solution does not exist in the extremal form.

\section{Discussion}
In this paper, we reconsidered generic solutions for partially
localized black asymptotically flat dyonic branes for three
particular values of the coupling constant
$a=0,\;a^2=n-1,\;a^2=3(n-1)$. For the first two the system separates
in terms of the Liouville equations, while in the third on can
construct an open three-dimensional Toda chain which generates the
desired solution. In all cases the solutions are four-parametric and
possess two curvature singularities (contrary to the singly charged
branes, for which the solutions have one singularity). All solutions
satisfy the cosmic censorship conjecture, i.e., they do not contain
naked singularities. Liouville solutions exist both in black and
extremal versions, the extremality being understood as the
degeneracy of the event horizon. In the extremal case the brane
world-volume  possesses the $ISO(p,1)$ isometry  and the solution
has to be localized (with the spherically symmetric transverse space
only). For the Toda solution we did not find a regular extremal
limit at all.

Meanwhile, somewhat different conclusions were made in the Ref.
\cite{Lu1996:la}, and we would like to clarify the situation here.
In this paper, only the isotropic localized dyonic branes were
considered ($B=D,\, E=0$ in our notation), and the solutions
obtained were not necessarily extremal (BPS). The solution of the
Ref. \cite{Lu1996:la}  can be found  imposing the following
conditions on the parameters of our most general solution (denoting
the parameters of \cite{Lu1996:la} by tilde): \bea
\lal n=k=\tilde{n},\qquad\tau=\tilde{\xi},\nn\\
\lal b_{1,2}=\tilde{\lambda}_{1,2},\qquad d_{0,1}=e_{0,1}=0,\qquad
\beta=4\tilde{k}, \eea for  $a=0,\;a^2=n-1$ \bea
\alpha=\tilde{k}\sqrt{8\tilde{n}},\qquad\tau_{1,2}=2\tilde{\alpha}_{1,2}
\eea for ${a^2=3(n-1)}$ \bea \tilde{n}=2,\;\alpha=4\tilde{k}, \qquad
B_1=\ln c_1c_2,\qquad B_2=\ln c_1. \eea It was checked in
\cite{Lu1996:la} that this solution satisfies the constraint
(\ref{ceqdp}) and the curvature scalar is finite on the event
horizon. However one can see that the Kretchmann scalar
(\ref{krech}) diverges there: \bea
K\sim\frac{\alpha^4}8\,(n-1)(n-2)(2n-3)\,\e^{-2\alpha\tau}. \eea
 Thus, the non-BPS dyonic branes with the isometry $ISO(p,1)$ found in the Ref.
\cite{Lu1996:la} are not regular on the horizon. In our analysis the
regularity of the horizon was ensured from the beginning by imposing
conditions of geodesic prolongation through the horizon (see the
section \ref{afrs}). In the non-extremal case (the non-degenerate
horizon) this gives an additional condition  on the horizon  for the
function $ \tilde{A} $ of the Ref. \cite{Lu1996:la}, namely,
$\dot{\tilde{A}}=0$ ($\dot{D}=0$ in our notation). From this
condition one finds \bea \tilde{k}=0. \eea Then in the Liouville
case one is led to the regular BPS brane.

For the Toda solution, as we have shown, the analysis of geodesics
excludes the possibility of the regular extremal limit. In terms
of the Ref. \cite{Lu1996:la}, the non-extremal solution with the
$ISO(p,1)$ symmetry must satisfy $\tilde{k}=0$, which condition
leads to $a=0$ in contradiction with the initially assumed value
of $a$.

Thus we conclude that in the case of the symmetry $ISO(p,1)$ the
only regular solutions are the standard BPS ones, and these exist
only in the Liouville cases  ($a=0,\;a^2=n-1$). In the Toda case
($a^2=3(n-1)$) regular solutions are necessarily black. Of course,
relaxing the condition of regularity (admitting naked
singularities) one finds larger classes of solutions.

The black dyon solution of the Ref. \cite{Duff:1996tb}   can be
obtained from our solution (\ref{lsolr}) imposing the following
conditions on the parameters
\bea \label{dbd}
\lal n=k,\qquad\xi_+=\tilde{k},\qquad\xi_-=0,\nn\\
\lal \xi_{1,2}=-\tilde{k}\sinh^2\tilde{\mu}_{1,2}. \eea Note that
our black solution is more general in the sense of the possibility
of the non-spherical transverse space (partial localization).  Note
also that in the extremal case the solution of the Ref.
\cite{Duff:1996tb} degenerates: when $\tilde{k}\rightarrow0$, we
find $\xi_-=\xi_+=\xi_1=\xi_2=0$. It is not necessarily so for our
general solution.

Here we restricted attention by the asymptotically flat dyonic
branes. Relaxing the asymptotic conditions, but still demanding the
absence of naked singularities, we are led to another possibility --
non-asymptotically flat dyonic branes with the linear dilaton
background at spatial infinity \cite{9}.

{\bf Acknowledgments.} The authors are grateful to Gerard Clement
for helpful discussions. This work was supported by RFBR grant
02-04-16949.


\begin{thebibliography}{99}

\bibitem{Lu1996:la}
H.~Lu and C.~N.~Pope, W.~Xu, ``Liouville and Toda Solutions in
M-theory'', {\it Mod. Phys. Lett.} {\bf A11}, 1785-1796 (1996);
hep-th/9604058.

\bibitem{2}V.~D.~Ivashchuk, V.~N.~Melnikov, ``P-brane black holes for
general intersections'', {\it Grav. Cosmol.} {\bf5}, 313-318 (1999); gr-qc/0002085.

\bibitem{ChGaGu02} Chiang-Mei Chen, Dmitri V. Gal'tsov, Michael Gutperle, ``S-brane
Solutions in Supergravity Theories'', {\it Phys.Rev.} {\bf D66}, 024043 (2002).

\bibitem{1} V.~D.~Ivashchuk, ``Composite fluxbranes with general intersections'',
{\it Class. Quant. Grav.} {\bf19}, 3033-3048 (2002); hep-th/0202022.

\bibitem{14}P. Brax, G. Mandal and Y. Oz, ``Supergravity description of
non-BPS branes'', {\it Phys. Rev. }{\bf D63},  064008 (2001).

\bibitem{141}
K.~A.~Bronnikov, V.~D.~Ivashchuk and V.~N.~Melnikov, ``The
Reissner-Nordstr\"om Problem for Intersecting Electric and
Magnetic p-Branes'', {\it Grav. Cosmol.} {\bf 3} 203 (1997);
gr-qc/9710054.

\bibitem{15}
V.~D.~Ivashchuk and V.~N.~Melnikov, ``Multidimensional classical
and quantum cosmology with intersecting
  p-branes'',{\it J. Math. Phys.}, {\bf 39} 2866 (1998); hep-th/9708157.

\bibitem{16}
V.~D.~Ivashchuk and V.~N.~Melnikov, ``Multidimensional classical
and quantum cosmology with intersecting
  p-branes'',{\it J. Math. Phys.}, {\bf 39} 2866 (1998); hep-th/9708157.

\bibitem{3} V.~D.~Ivashchuk, V.~N.~Melnikov, ``Black hole p-brane solutions
for general intersection rules'', {\it Grav. Cosmol.} {\bf6},
27-40 (2000); hep-th/9910041.

\bibitem{4} M.~A.~Grebeniuk, V.~D.~Ivashchuk, ``Sigma-model Solutions and
Intersecting p-Branes Related to Lie Algebras'', {\it Phys. Lett.}
{\bf B442}, 125-135 (1998); hep-th/9805113.

\bibitem{10}Y.~G.~Miao and N.~Ohta, ``Complete Intersecting Non-Extreme p-Branes''
{\it Phys.\ Lett.\ } {\bf B594}, 218 (2004); hep-th/0404082.

\bibitem{GaLeCl04}  D. Gal'tsov, J. Lemos and G. Cl\'ement,
``Supergravity $p$-brane reexamined: extra parameters, uniqueness
and topological censorship'', {\it Phys. Rev. }{\bf D70}, 024011
(2004); hep-th/0403112.

\bibitem{Gal'tsov:2005vf}
D.~Gal'tsov, S.~Klevtsov, D.~Orlov, and G.~Clement, ``More on
general p-brane solutions'', {\it Int. J. Mod. Phys. A} in
press;hep-th/0508070.

\bibitem{5}  Miguel~S.~Costa, ``Composite M-branes'', {\it Nucl.Phys.}
{\bf B490}, 202-216 (1997); hep-th/9609181.

\bibitem{6} J.~M.~Izquierdo, N.~D.~Lambert, G.~Papadopoulos,
P.~K.~Townsend, ``Dyonic Membranes'', {\it Nucl.Phys.} {\bf B460},
560-578 (1996); hep-th/9508177.

\bibitem{ClGa04}
G.~Cl\'ement, D.~Gal'tsov and C.~Leygnac, ``Black branes on the
linear dilaton background'', {\it Phys. Rev. }{\bf D71}, 084014
(2005), hep-th/0412321.

\bibitem{Gavrilov:1994mc}
V.~R.~Gavrilov, V.~D.~Ivashchuk, V.~N.~Melnikov,
``Multidimensional cosmology with multicomponent perfect fluid and
Toda lattices'', gr-qc/9407019.

\bibitem{Olshanetsky:1979} M.~A.~Olshanetsky, A.~M.~Perelomov,
``Explicit Solutions of Classical Generalized Toda Models``, {\it
Invent. Math.} {\bf 54}, 261 (1979).

\bibitem{12}
B.Kostant, {\it Adv. in Math.} {\bf 34}, 195 (1979).

\bibitem{13}
A.~N.~Leznov and M.~V.~Saveliev, ``Group Theoretical Methods for Integration
of Nonlinear Dynamical Systems'', Nauka, Moscow, 1985.

\bibitem{Duff:1996tb}
M.~J.~Duff, H.~Lu and C.~N.~Pope, ``The Black Branes of
M-theory'', {\it Phys. Lett.}  {\bf B382}, 73 (1996); hep-th/9604052.

\bibitem{9} G.~Cl\'ement, D.~Gal'tsov, C.~Leygnac and D.~Orlov,
``Dyonic  branes  and linear dilaton background'', hep-th/0512013.

\end{thebibliography}
\end{document}